\documentclass[12pt]{article}
\usepackage{amsmath,amssymb,color}
\usepackage{cite,epsf}
\usepackage{pstricks}
\usepackage{color}

\usepackage{graphicx,array} 
\newcommand{\be}{\begin{equation}}
\newcommand{\ee}{\end{equation}}
\newcommand{\beq}{\begin{equation}}
\newcommand{\eeq}{\end{equation}}
\newcommand{\bea}{\begin{eqnarray}}
\newcommand{\eea}{\end{eqnarray}}

\newcommand{\ba}{\begin{eqnarray}}
\newcommand{\ea}{\end{eqnarray}}

\def\sin{\mbox{sin}}
\def\cos{\mbox{cos}}
\def\cot{\mbox{cot}}
\def\log{\mbox{log}}

\begin{document}

\begin{titlepage}
\vspace{10pt}
\hfill
{\large\bf HU-EP-16/27}
\vspace{20mm}
\begin{center}

{\Large\bf  On the logarithmic
divergent part of entanglement entropy, smooth versus singular regions   \\[2mm]

}

\vspace{45pt}

{\large Harald Dorn 
{\footnote{dorn@physik.hu-berlin.de
 }}}
\\[15mm]
{\it\ Institut f\"ur Physik und IRIS Adlershof, 
Humboldt-Universit\"at zu Berlin,}\\
{\it Zum Gro{\ss}en Windkanal 6, D-12489 Berlin, Germany}\\[4mm]

\vspace{20pt}

\end{center}
\vspace{10pt}
\vspace{40pt}

\centerline{{\bf{Abstract}}}
\vspace*{5mm}
\noindent
The entanglement entropy for smooth regions $\cal A$ has a logarithmic divergent
contribution with a shape dependent coefficient and that for regions
with conical singularities an additional $\log ^2$ term.
Comparing the coefficient of this extra term, obtained by direct holographic
calculation for an infinite cone, with the corresponding limiting case for 
the shape dependent coefficient for a regularised cone, a mismatch by a 
factor two has been observed in the literature. We discuss several aspects 
of this issue. In particular a regularisation of  $\cal A$, intrinsically 
delivered 
by the holographic picture, is proposed and applied to an example  
of a compact region with two conical singularities.
Finally, the mismatch is removed in all 
studied regularisations of $\cal A$, if equal scale ratios are chosen for the limiting procedure. 

\vspace*{4mm}
\noindent

\vspace*{5mm}
\noindent

\end{titlepage}
\newpage


\section{Introduction}

The entanglement entropy
for compact three-dimensional regions $\cal A$ with a smooth boundary $\partial\cal A$ in $(3+1)$-dimensional conformal quantum field theories has 
UV-divergent contributions. The leading one is quadratic $\propto 1/\epsilon^2$
with a coefficient proportional to the area of $\partial\cal A $. The nextleading
term is logarithmic $\propto\log\epsilon$ with a shape dependent coefficient derived by Solodukhin in \cite{Solodukhin:2008dh}.   
His formula for the holographic evaluation in the case of strong coupling ${\cal N}=4$ SYM and static regions $\cal A$ in $\mathbb{R}^{(3,1)}$ is \cite{Solodukhin:2008dh}
\beq
S({\cal A})=\frac{1}{4G_N^{(5)}}\left ( \frac{A(\partial{\cal A})}{2\epsilon^2}~+~K~\log \epsilon ~+~{\cal O}(1)~\right )\label{solo1}
\eeq
with
\beq
K~=~\frac{1}{8}~\int_{\partial{\cal A}}~k^2~ \sqrt{\mbox{det}~g}~d^2z ~,\label{K}
\eeq
where $G_N^{(5)}$ is the 5-dim. Newton constant, $g$ the induced metric
on $\partial{\cal A} \subset \mathbb{R}^{(3,1)}$ and $k$ the trace of its second fundamental form. Obviously, the coefficient $K$ becomes divergent if the surface
$\partial{\cal A} $ develops singularities. This is in correspondence
to the appearance of a $\log^2\epsilon$ term in the direct holographic
calculation for regions $\cal A$ with conical singularities via the
Ryu-Takayanagi formula \cite{Ryu:2006bv,Ryu:2006ef}
\beq
S({\cal A})=\frac{V(\gamma_{\cal A})}{4G_N^{(5)}}~.
\eeq
There $V(\gamma_{\cal A})$ denotes the volume of the minimal spatial $3$-dimensional submanifold $\gamma_{\cal A}\subset AdS_{5}$, approaching the boundary $\partial {\cal A}$ on the boundary of $AdS_5$.

The calculation of the regularised volume of $\gamma _{\cal A}$  with UV cutoff $r>\epsilon$ and IR cutoff $\rho <l$ for the case, 
\footnote{$~~r$ is the Poincar{\' e} coordinate pointing into the interior of $AdS_5$ and $\rho$ is the Euclidean distance from the tip of the cone.}
where $\cal A$ is an infinite cone with opening angle $2\Omega$,
 yields \cite{Klebanov:2012yf, Myers:2012vs,Dorn:2016eai} 
\beq
S_{\epsilon,l}~=~=\frac{1}{4G_N^{(5)}}\left (\frac{\pi ~ \sin\Omega}{2}\frac{l^2}{\epsilon^2}~-~\frac{\pi ~\cos\Omega~\cot\Omega}{8}~\log ^2\frac{\epsilon}{l}~+~\cal O(\log\epsilon)\right )~.\label{direct}
\eeq
The authors of  \cite{Klebanov:2012yf, Myers:2012vs} have raised the question of
how the coefficient of the above $\log^2$ term can be obtained out
of Solodukhin's formula and found agreement up to a mismatch by a numerical factor 2. An analog observation in (5+1) dimensions was made in \cite{Safdi:2012sn}. Furthermore, the mismatch factor 2 was observed also for certain perturbed spheres in even
dimensional CFT's \cite{ Bueno:2015lza}.  
 
Let us sketch the line of reasoning in \cite{Klebanov:2012yf, Myers:2012vs} and  parameterise $\partial {\cal A}$, the boundary of the cone,  by the coordinates $\rho,\varphi$
\beq 
x_1=\rho~ \sin\Omega~\cos\varphi~,~~x_2=\rho ~\sin\Omega~\sin\varphi~,x_3=\rho~\cos\Omega ~.
\eeq
Then the trace of the second fundamental form is
\beq
k~=~\frac{\cot \Omega}{\rho }~,
\eeq
and the square root of the induced metric
\beq
\sqrt{\mbox{det}~g}~=~\rho ~\sin\Omega~.
\eeq
For a sphere of radius $R$ the corresponding quantities are (use spherical coordinates
$\vartheta,\varphi$)
\beq
k_{\mbox{\scriptsize sphere}}~=~\frac{2}{R}~,~~~\sqrt{\mbox{det}~g_{\mbox{\scriptsize sphere}}}~=~R^2~\sin\vartheta ~.
\eeq
If one regularises the singular geometry at the tip of the cone by fitting a piece
of a small sphere, this piece does {\it not} contribute to the divergence
of $K$ since the dependence on its radius cancels in the integral \eqref{K}. Therefore we have
\beq
K~=~\frac{2\pi}{8}~\int _{\rho_{\mbox{\scriptsize min}}}^l\cot^2\Omega~\sin\Omega~\frac{d\rho}{\rho}~+~{\cal O}(1)~=~-~\frac{\pi~\cos\Omega~\cot\Omega}{4}~\log\frac{\rho_{\mbox{\scriptsize min}}}{l}~+~{\cal O}(1)~.
\eeq
By the natural identification of $\rho_{\mbox{\scriptsize min}}$ with $\epsilon$
one gets $K\log\epsilon=-~\frac{\pi~\cos\Omega~\cot\Omega}{4}~\log^2\epsilon~+~\dots$
and the mismatch by a factor of 2 relative to the direct holographic calculation \eqref{direct} as
observed in \cite{Klebanov:2012yf, Myers:2012vs}.

At this point it  is tempting to suspect IR/UV mixing under conformal transformations for this mismatch. A corresponding
argument could start as follows. 
By the natural choice $l=1/\rho_{\mbox{\scriptsize min}}$ we would arrive at
\beq
K~=~-~\frac{\pi~\cos\Omega~\cot\Omega}{2}~\log\rho_{\mbox{\scriptsize min}}~+~{\cal O}(1)~.
\eeq
After $\rho_{\mbox{\scriptsize min}}=\epsilon$ this agrees perfectly with
what one gets from the direct calculation \eqref{direct}  after  $l=1/\epsilon$
 (note: $\log^2\epsilon^2=4\log^2\epsilon $). However, this is a doubtful reasoning, since it immediately breaks down
if one uses the dimensionless quotient $\epsilon/l$ as the argument of the log in eq.\eqref{solo1}, too. This again  would bring back the factor 2. 

What remains from this aside is, that for a clean discussion of the behaviour of Solodukhins formula in the limit of singular boundaries $\partial {\cal A}$, we have 
to rely on its use for compact regions $ {\cal A} $. 
\section{Coefficient of logarithmic divergence for 
banana shaped regions with rounded tips }
As an example for a compact region  $ {\cal A} $ with two conical singularities we take
a banana shaped region as studied in our paper \cite{Dorn:2016eai}. In a first
attempt, for the regularisation we apply the technique used in the previous
section: replacement of the conical tips by suitable fitted parts of small
spheres. The boundary $\partial{\cal A} $ is given by
\bea
x_1(\rho,\varphi)&=& \frac{\rho ~\cos\alpha ~\sin\Omega~\cos\varphi + 
    \rho ~\sin\alpha ~\cos\Omega}{q^2 + \rho^2 + 
    2 q \rho~\hat w(\varphi)}~,\nonumber\\[2mm]
x_2(\rho, \varphi) &= &
 \frac{\rho~ \sin\Omega~ \sin\varphi}{q^2 + \rho^2 + 
     2 q \rho~\hat w(\varphi)}~,\nonumber\\[2mm]
x_3(\rho, \varphi) &=& \frac{q + \rho ~\hat w(\varphi)}{q^2 + \rho^2 + 
     2 q \rho~\hat w(\varphi)}~,\label{banana}
\eea
with $0\leq\rho<\infty,~0\leq \varphi<2\pi $ and
\beq
\hat w(\varphi)~=~\cos\alpha ~\cos\Omega - \sin\alpha ~\sin\Omega~ \cos\varphi~.
\eeq 
$2\Omega$ is the opening angle of the conical tips, $\alpha$ is the angle between 
its axis\footnote{For $\alpha >0$ this axis is the piece of a circle.} and the straight line
connecting the tips, $1/q$ is the distance between the tips.

The induced metric on $\partial{\cal A} $ is
\bea
g_{\rho\rho}&=&\frac{1}{\big (q^2 + \rho^2 + 2 q\rho~\hat w(\varphi)\big )^2}~,~~~g_{\rho\varphi}~=~0~,\nonumber\\
g_{\varphi\varphi}&=&\frac{
\rho^2~ \sin^2\Omega}{\big (q^2 + \rho^2 + 2 q\rho~\hat w(\varphi)\big )^2}~.
\eea
The second fundamental form turns out as
\bea
k_{\rho\rho}&=& \frac{
  2 q ~( \sin\alpha~\cos\Omega~\cos\varphi + \cos\alpha~ \sin\Omega)}{\big (q^2 + \rho^2 + 
    2 q \rho ~\hat w(\varphi)\big )^2}~,~~~k_{\rho\varphi}~=~0~,
\nonumber\\
k_{\varphi\varphi}&=&\frac{\rho~\sin\Omega~ \big (2 q \rho ~\cos\alpha + (q^2 + \rho^2) 
\cos\Omega\big )}{\big (q^2 + \rho^2 + 
    2 q \rho~\hat w(\varphi)\big )^2}~,
\eea
and its trace is
\beq
k:=g^{mn}k_{mn}=~2q(\sin\alpha~\cos\Omega~\cos\varphi+\cos\alpha~\sin\Omega)~+~\frac{2q\rho~\cos\alpha+(q^2+\rho^2)\cos\Omega}{\rho~\sin\Omega}~.
\eeq
The integrand in Solodukhin's formula \eqref{K} behaves for $\rho\rightarrow 0$ as
\beq
k^2~ \sqrt{\mbox{det}~g}~=~\frac{\cos\Omega~\cot\Omega}{\rho}~+~{\cal O}(1)
\eeq
and for  $\rho\rightarrow \infty$ as
\beq
k^2~ \sqrt{\mbox{det}~g}~=~\frac{\cos\Omega~\cot\Omega}{\rho}~+~{\cal O}(1/\rho^2)~.
\eeq
As discussed in the previous section, the regularising spherical pieces do not 
contribute to the divergence.
Performing the $\varphi$-integration and cutting the logarithmic divergent $\rho$-integration at  $\rho_{\mbox{\scriptsize min}}$ and $\rho_{\mbox{\scriptsize max}}$ we get
\beq
K~=~\frac{2\pi}{8}~\cos\Omega~\cot\Omega~(-\log\rho_{\mbox{\scriptsize min}}~+~\log\rho_{\mbox{\scriptsize max}})~+~\dots ~,
\eeq
where the dots stands for terms staying finite if one removes the cutoffs. With $\rho_{\mbox{\scriptsize min}}=1/\rho_{\mbox{\scriptsize max}}=\epsilon $ this yields
\beq
K\log\epsilon~=~-\frac{\pi}{2}~\cos\Omega~\cot\Omega~\log^2\epsilon ~+~{\cal O}(\log\epsilon)~.
\eeq
Comparing this with our result \cite{Dorn:2016eai} for the direct holographic calculation in the case of unregularised conical tips, one again gets a mismatch by a factor 2.

From this example we can conclude that the origin of the mismatch is {\it not}
related to the IR/UV issue. Instead it has to be located in the use of different
limiting procedures. In the direct holographic calculation one uses only {\it one}
cutoff (Poincar{\' e} coordinate $r>\epsilon$) for the volume of the minimal submanifold $\gamma_{\cal A}$ related to a region $\cal A$ with conical singularities. On the other side at first the same holographic recipe is applied for a smoothed $\cal A$
obtained by rounding the conical singularities. This rounding introduces further
independent regularisation parameters ($\rho_{\mbox{\scriptsize min}},\rho_{\mbox{\scriptsize max}}$). Relating them to $\epsilon$ as above  sounds natural but, taken seriously, is an ambiguous procedure. Note also that e.g. $\rho_{\mbox{\scriptsize min}}=1/\rho_{\mbox{\scriptsize max}}=\epsilon ^{1/2}$ would remove the unwanted mismatch factor \nolinebreak 2. We will come back to this point in the conclusion section. 

But before we would like to explore another option, to replace the handmade
regularisation of $\partial{\cal A}$ for the use in Solodukhin's formula \eqref{K} by one which is delivered
by the holographic recipe itself. Let us consider the  minimal submanifold $\gamma_{\cal A}$ needed for the treatment of a singular region $\cal A$. Then for use in \eqref{K} 
we take its intersection with the hyperplane $r=\epsilon $ as our regularised
version of $\partial\cal A$.\footnote{Remember $\partial\cal A$ is the intersection with the boundary of $AdS$ at $r=0$. } This procedure we will demonstrate in the next section with its application to lemon shaped regions. 
\section{Coefficient of log$\epsilon$ for 
lemon shaped regions with a holographically induced regularisation}

In \cite{Dorn:2016eai} the minimal submanifold $\gamma_{\cal A}$ in Euclidean $AdS_4$ \footnote{We discuss static regions $\cal A$, therefore time is frozen.}, whose volume
up to the factor $\frac{1}{4G_N^{(5)}}$ determines the holographic entanglement entropy of a banana shaped region \eqref{banana}, has been obtained
\bea
x_1&=&\frac{\rho ~(\cos\alpha~\sin\vartheta~\cos\varphi~+~\sin\alpha~\cos\vartheta)}{q^2~+~\rho^2(1+h^2(\vartheta)) ~+2q \rho ~w(\vartheta,\varphi)},\nonumber\\[2mm]
x_2&=&\frac{\rho~\sin\vartheta~\sin\varphi}{q^2~+~\rho^2(1+h^2(\vartheta)) ~+2q \rho ~w(\vartheta,\varphi)},\nonumber\\[2mm]
x_3&=&\frac{q~+~\rho~w(\alpha,\vartheta,\varphi)}{q^2~+~\rho^2(1+h^2(\vartheta)) ~+2q \rho ~w(\vartheta,\varphi)},\nonumber\\[2mm]
r&=&\frac{\rho ~h(\vartheta)}{q^2~+~\rho^2(1+h^2(\vartheta)) ~+2q \rho ~w(\vartheta,\varphi)}~,\label{surface}
\eea
with $w(\vartheta,\varphi)~=~\cos\alpha~\cos\vartheta~-~\sin\alpha~\sin\vartheta~\cos\varphi$ and  $h(\vartheta )$ the solution of the differential equation
\bea
\big (~\ddot h (h+h^3)+\dot h^2(3+h^2)+3+5h^2+2h^4~\big )&\sin\vartheta & \nonumber\\[2mm]
+~h\dot h(1+h^2+\dot h^2)&\cos\vartheta &=~0~\label{eom}
\eea
and the boundary condition $h(\Omega)=0,~~h(0)=h_0(\Omega)>0$.

Here $\rho,\vartheta,\varphi$ are coordinates and $q,\Omega,\alpha$ parameters
fixing the geometry of the banana shaped region as described in the previous section. The regularised
volume of  this $\gamma_{\cal A}$ has been calculated up to terms vanishing for $\epsilon\rightarrow 0$.
We show here only the $\log^2\epsilon$ term
\beq
V_{\epsilon}~=~\dots ~-\frac{\pi~\cos\Omega~\cot\Omega}{4}~\log^2(q\epsilon)~+~\dots
~.\eeq

As announced above, for a regularised version of ${\cal A}$ we take the intersection 
of this submanifold with the hyperplane $r=\epsilon$, see fig.1. Then $\partial {\cal A}_{\mbox{\scriptsize reg}}$
is parameterised by the coordinates $\vartheta$ and $\varphi$ via \footnote{For simplicity we consider only the symmetric case $\alpha =0$, which corresponds to a kind of lemon shape. Then $w=\cos\vartheta$.}
\bea
x_1&=&\frac{\rho_{\epsilon}^{\pm}(\vartheta )~\sin\vartheta~\cos\varphi}{q^2+(\rho_{\epsilon}^{\pm})^2(1+h^2(\vartheta))+2q\rho_{\epsilon}^{\pm}~\cos\vartheta}~,\nonumber\\
x_2&=&\frac{\rho_{\epsilon}^{\pm}(\vartheta )~\sin\vartheta~\sin\varphi}{q^2+(\rho_{\epsilon}^{\pm})^2(1+h^2(\vartheta))+2q\rho_{\epsilon}^{\pm}~\cos\vartheta}~,\nonumber\\
x_3&=&\frac{q+\rho_{\epsilon}^{\pm}(\vartheta )~\cos\vartheta~}{q^2+(\rho_{\epsilon}^{\pm})^2(1+h^2(\vartheta))+2q\rho_{\epsilon}^{\pm}~\cos\vartheta}~,
\eea
where $\rho_{\epsilon}^{\pm}$ are the two roots of the equation $r=\epsilon$, i.e.
\beq
\rho^{\pm}_{\epsilon}(\vartheta )~=~\frac{h(\vartheta)-2q\epsilon~\cos\vartheta \pm\sqrt{(h-2q\epsilon ~\cos\vartheta)^2-4\epsilon ^2q^2(1+h^2)}}{2\epsilon~(1+h^2)}~.\label{rhopm}
\eeq
Each of these two roots is responsible for the description of a half
of the regularised boundary of our lemon shaped region. 
\begin{figure}[h!]
 \centering
 \includegraphics[width=13cm]{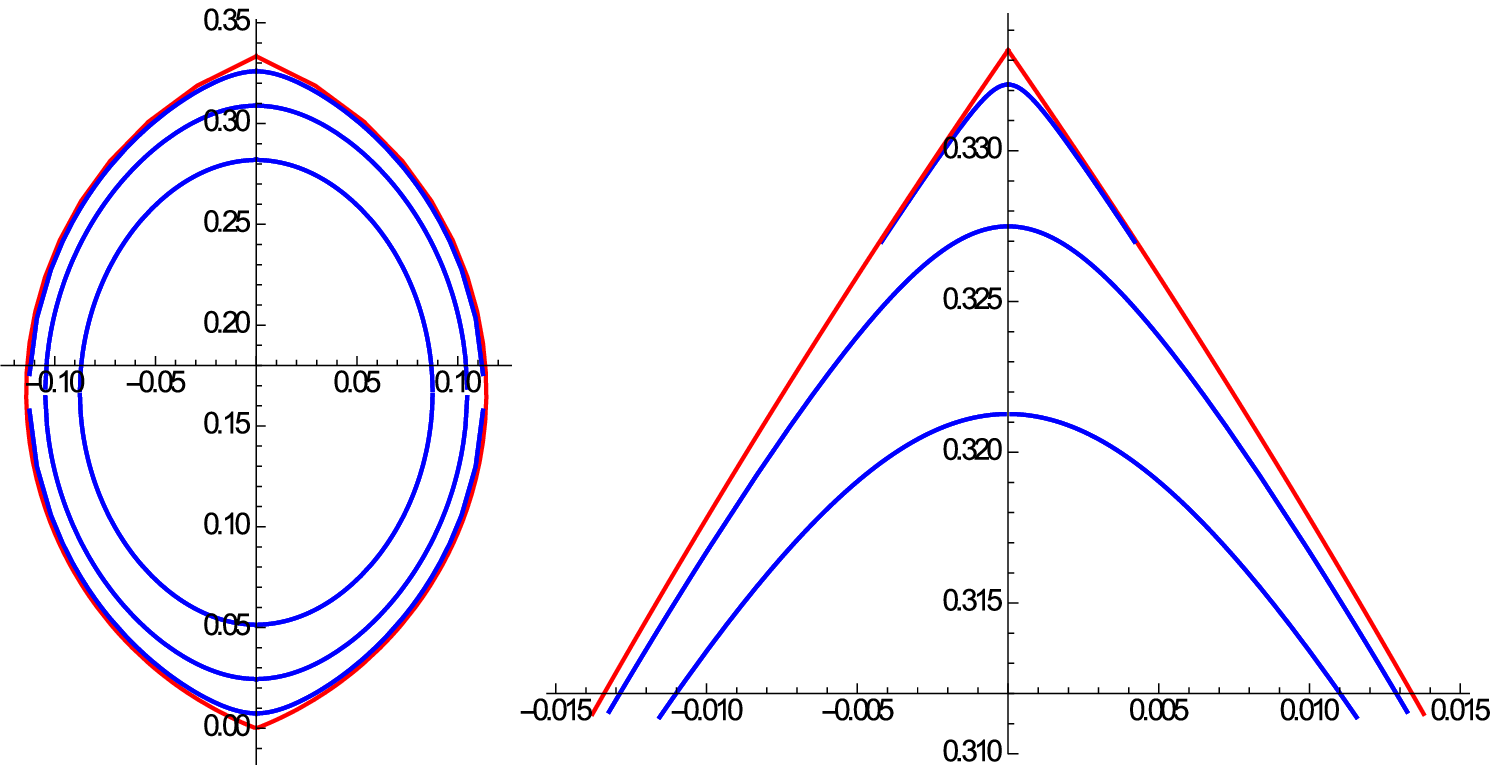} 
\caption{\it Axial cut of  (regularised) lemons=symmetric bananas ($\alpha =0$) with $q=3$.\newline $~~~~~~~~~~~~$ In red the original lemons, in blue regularised versions.\newline  On the left: the full picture for $\Omega\approx 1.2$ and $\epsilon=0.02, 0.05, 0.08~.$ \newline
On the right: zoom into the upper tip for $\Omega\approx 0.6$ and $\epsilon =0.001, 0.005, 0.01$.  } 
\label{reg-ban}
\end{figure}

The induced metric is ($\rho$ stands for $\rho_{\epsilon}^{\pm}(\vartheta)$, $h$ for $h(\vartheta)$ and the dot for $d/d\vartheta$ ) 
\bea
g_{\vartheta\vartheta}&=&\frac{1}{\big (q^2 + 
   2 q \rho~ \cos\vartheta + (1 + h^2) \rho^2\big )^4}~\Big\{\Big (\rho \big((-q^2 + (1 + h^2)\rho^2) \sin\vartheta \nonumber\\
&&+ 
        2 h\dot h\rho (q +\rho ~\cos\vartheta)  \big ) + \dot \rho \big (q^2 \cos\vartheta + \rho (1 + h^2) (2 q +\rho~ \cos\vartheta )\big ) \Big)^2\nonumber\\
&& + \Big (\rho \big (q^2 \cos\vartheta + 2 q \rho + 
        \rho^2 ((1+h^2)~\cos\vartheta  - 
           2 h\dot h~ \sin\vartheta)\big )\nonumber\\
&& + \dot\rho\big (q^2 - (1 + 
           h^2) \rho^2\big ) ~\sin\vartheta \Big )^2\Big\}~,\\
g_{\varphi\varphi}&=&\frac{
 \rho^2~\sin^2\vartheta}{(q^2 + 
   2 q\rho~ \cos\vartheta  + (1 + h^2) \rho^2)^2}~,~~~~~~~~g_{\vartheta\varphi}~=~0~.
\eea 
Inserting \eqref{rhopm} one gets for the determinant of the induced metric the same
expression {\it both} for the plus and the minus variant
\bea
\mbox{det}g(\vartheta)&=&\frac{\epsilon ^4 \sin ^2\vartheta}{h^4 \big (4 \epsilon qh~ \cos\vartheta  - (1 - 4 \epsilon^2 q^2) h^2 + 
     4 \epsilon^2 q^2 \sin^2\vartheta \big )}\nonumber\\
&&\cdot\Big ( 4 \epsilon q h~\cos\vartheta 
 - 4 \epsilon q\dot h~ \sin\vartheta  - 
     4 \epsilon^2 q^2 h\dot h~ \sin(2\vartheta)\nonumber\\
&&~~ +     2 \epsilon^2 q^2 (h^2 - \dot h^2) ~ \cos(
       2 \vartheta)  
- (1 -  2 \epsilon^2 q^2) (h^2 + \dot h^2)\Big )~,
\eea
which implies for the $\epsilon$-expansion of its square root
\beq
\sqrt{\mbox{det}g(\vartheta)}~=~\frac{\sin\vartheta~\sqrt{h^2+\dot h^2}}{(h(\vartheta))^3}~\epsilon ^2~+~\frac{2q~\sin\vartheta~(h\dot h~\sin\vartheta+\dot h^2\cos\vartheta)}{h^4\sqrt{h^2+\dot h^2}}~\epsilon^3~+~{\cal O}(\epsilon^4)~.\label{sqrtgexpanded}
\eeq 
The trace of the second fundamental form \footnote{Note $k_{\vartheta\varphi} =0$, $\vec n(\vartheta,\varphi)$ denotes the normal vector fixed up to a sign by $\vec n^2=1,~\vec n\partial_{\vartheta} \vec x=\vec n\partial _{\varphi}\vec x=0$. Due to the symmetry of our lemon shaped surface, $k$ depends on $\vartheta$ only.}
\beq
k(\vartheta)=g^{\vartheta\vartheta}~\vec n~\frac{\partial^2\vec x}{\partial \vartheta^2}~+~g^{\varphi\varphi}~\vec n~\frac{\partial^2\vec x}{\partial \varphi^2}
\eeq
differs in its plus/minus variant after inserting $\rho_{\epsilon}^{\pm}$ from \eqref{rhopm}.
Expanding the arising longer expressions in $\epsilon$ one gets
\beq
k^{\pm}(\vartheta)~=~\frac{h\big (2h^3+h\dot h^2\pm\dot h^3\cot\vartheta \pm h^2(\dot h~\cot\vartheta
+ \ddot h)\big )}{(h^2+\dot h^2)^{3/2}}~\frac{1}{\epsilon} ~ +~{\cal O} (1)~
\eeq
and then with \eqref{sqrtgexpanded} for the integrand in Solodukhins formula \eqref{K} 
\beq
\sqrt{\mbox{det}g(\vartheta)}~(k^{\pm}(\vartheta))^2~=~\frac{\big (2h^3+h\dot h^2\pm\dot h^3\cot\vartheta \pm h^2(\dot h~\cot\vartheta
+ \ddot h)\big )^2~\sin\vartheta}{h(h^2+\dot h^2)^{5/2}}~+~{\cal O}(\epsilon)~.
\eeq
From our paper \cite{Dorn:2016eai} we know for $\vartheta=\Omega-\delta~,~~~\delta
\rightarrow 0$
\beq
h(\vartheta)~=~2~ (\tan\Omega)^{1/2} \delta ^{1/2}~+~{\cal O}(\delta^{3/2}\log\delta)~,\label{hasympt}
\eeq
what implies
\bea
\sqrt{\mbox{det}g}~(k^{+})^2&=&\frac{\cos\Omega~\cot\Omega}{2\delta}~+~\cos\Omega+\frac{1}{2}~\cos\Omega~\cot^2\Omega~+~{\cal O}(\delta)~+~{\cal O}(\epsilon)~,\\
\sqrt{\mbox{det}g}~(k^{-})^2&=&\frac{\cos\Omega~\cot\Omega}{2\delta}~+~5~\cos\Omega+\frac{1}{2}~\cos\Omega~\cot^2\Omega~+~{\cal O}(\delta)~+~{\cal O}(\epsilon)~.\nonumber
\eea
The not explicitly shown ${\cal O}(\epsilon)$ terms behave as $\delta^{-3/2}$.

With this estimates we get from \eqref{K}
\bea
K&=&\frac{1}{8}\int_0^{2\pi}d\varphi\int_0^{\Omega-\delta_{{\mbox{\scriptsize min}}}}
d\vartheta \sqrt{\mbox{det}g(\vartheta)}~\big ((k^{+}(\vartheta))^2+(k^{-}(\vartheta))^2\big )\nonumber\\[2mm]
&=&\frac{\pi}{4}~\frac{\cos\Omega~\cot\Omega}{2}~(-2~\log\delta_{{\mbox{\scriptsize min}}})~+~{\cal O}(1)~+~\delta_{{\mbox{\scriptsize min}}}^{-1/2}\cdot{\cal O}(\epsilon)~. 
\eea
In the regularisation chosen in this section the upper boundary of the $\vartheta$ integration (lower bd. for $\delta $) is fixed by the vanishing of the expression under the square root in \eqref{rhopm}. This corresponds to fitting together the two halves of the regularised $\partial{\cal A}$. In this condition $\epsilon\rightarrow 0$ implies $h\rightarrow 0,~\vartheta\rightarrow\Omega~$. Therefore, with \eqref{hasympt} we get
\beq
\delta_{{\mbox{\scriptsize min}}}~=~q^2\cot^2\Omega~(1+\cos\Omega)^2~\epsilon^2~+~{\cal O}(\epsilon^3)~.
\eeq
Now we see two facts: At first, due to its singular $\delta ^{-3/2}$-behaviour,
the integration of the ${\cal O}(\epsilon)$ term of the integrand is not vanishing
for $\epsilon\rightarrow 0$, but it does not diverge. At second, due to $\delta_{{\mbox{\scriptsize min}}}\propto\epsilon^2$
\beq
K~\log\epsilon~=~-\frac{\pi}{2}~\cos\Omega~\cot\Omega~\log^2\epsilon ~+~{\cal O}(\log\epsilon)~.
\eeq
This is the same as in the previous section, i.e. the mismatch factor 2 is back.  
\section{Conclusions}
We have found the mismatch factor 2, observed in \cite{Klebanov:2012yf, Myers:2012vs}
for the infinite cone, also for prototypical compact regions with two conical
singularities. It appeared in section 2 using for Solodukhin's formula a handmade
regularisation of  $\partial {\cal A}$, independent of the holographic cut-off procedure. And it reappeared in section 3 using a regularisation of $\partial {\cal A}$,
delivered in a natural way by the holographic cut-off procedure itself.  

Due to this robustness it is time to answer the question of why this mismatch factor 
in the so far presented calculations is always just equal to 2.

The direct holographic calculation for $\partial {\cal A}$ depends on {\it one} cut-off parameter $\epsilon\rightarrow 0$. We compare it with $\epsilon\rightarrow 0$
for regularised  $\partial {\cal A}$ and the {\it subsequent} limit of removal
of the  $\partial {\cal A}$-regularisation. As usual a priori it is open, whether
the two different limits yield the same result.

Let us denote by $\epsilon '$ the parameter controlling the regularisation of $\partial {\cal A}$. In section \nolinebreak 2 we had $\epsilon '= \rho_{\mbox{\scriptsize min}}$. In
section 3 we used the intersection of the minimal submanifold $\gamma_{\cal A}$ with
the $AdS$-hyperplane $r=\epsilon$, but we could have done it also with another
hyperplane $r=\epsilon '$. Then the universal result for both types of regularisation for  $\partial {\cal A}$ is
\beq
K(\epsilon ')~\log\epsilon~=~-\frac{\pi}{2}~\cos\Omega~\cot\Omega~\log\epsilon '~\log\epsilon ~+~\dots ~.\label{universal}
\eeq
Instead of putting $\epsilon '=\epsilon$ we better should require that $\epsilon$
goes faster to zero than $\epsilon '$. Only in this manner we can keep contact with 
the appropriate order of limits,
\beq
\mbox{first}~~~ \epsilon \rightarrow 0 ~,~~ \mbox{subsequently}~~~ \epsilon '\rightarrow 0~.\label{order}
\eeq
 Then with $\epsilon '=\epsilon ^{\beta} ~,~~0<\beta<1$ we get
\beq
K~\log\epsilon~=~-\beta~\frac{\pi}{2}~\cos\Omega~\cot\Omega~\log^2\epsilon~+~\dots ~.
\label{ambiguity}
\eeq
The choice $\beta=1/2$ yields complete agreement with the direct holographic
calculation in \cite{Dorn:2016eai}. 

Instead stopping at this point with an ambiguity parametrised by the factor $\beta$,
one should stress that  $\beta=1/2$ is distinguished not only by the a posteriori
fit to the direct calculation.\footnote{$\beta=1/2$, justified a posteriori, was also discussed in \cite{Bueno:2015lza} for still another regularisation. } Remarkably, it also corresponds just to the 
choice of equal scale ratios to mimic the order of limits \eqref{order} in a one parameter set up :$~~\epsilon'/1=\epsilon/\epsilon '~$, i.e. the ratio of the scale for regularising the conical singularity to a constant
equals the ratio of the scale for approaching the $AdS$ boundary to the 
scale for regularising the cone.\\ 

Altogether the outcome of our study can be summarised as follows.  
Based on a naive identification of the two regularisation scales
($\epsilon=\epsilon '$),
the comparison of the factor for the logarithmic divergence, given 
by Solodukhin's
formula, in the limit where $\partial{\cal A}$ develops a conical 
singularity with the direct holographic calculation for the singular  
$\partial{\cal A}$ yields agreement up to a numerical mismatch factor of 
2 \cite{Klebanov:2012yf,Myers:2012vs,Safdi:2012sn,Bueno:2015lza}. In general the results of different limiting procedures
can disagree. Therefore, the fact that in the case under discussion the geometrical
structures on both sides are the same, and the only discrepancy is a numerical
factor, is already a remarkable result. We have shown that this mismatch factor 2 is robust with respect
to the choice of regularisations and, treating compact regions, has
nothing to do with the infrared issue for infinite cones. Furthermore,
we pinned down the value of the numerical mismatch factor to the implementation of the order of
limits \eqref{order}  in the relation of $\epsilon$ to $\epsilon '$. This order is a
constitutive ingredient of the path via Solodukhin's formula \eqref{K}. Given the universal formula \eqref{universal}, an absence of the mismatch factor for
the naive choice  would be even confusing, since 
$\epsilon=\epsilon '$ in no respect makes contact with the order of limits \eqref{order}. The most natural way to mimic this order is realised by the equal scale ratio, i.e. $\beta =1/2$. Then the absence of any mismatch is a consequence.
\newpage 

\end{document}